\documentclass[11pt,fleqn]{article}
\usepackage[scale={.83,.83}]{geometry}
\usepackage{amsmath,amssymb,amsfonts,dcolumn}
\usepackage[margin=-5pt]{caption}
\usepackage{axodraw,graphicx,wrapfig}
\usepackage[hypertex]{hyperref}

\newcolumntype{d}{D{.}{.}{0}}

\newlength{\textlength}
\newlength{\overlinelength}
\newcommand{\ovl}[2][.55]{\settowidth{\textlength}{$#2$}
  \setlength{\overlinelength}{0.1pt}
  \addtolength{\overlinelength}{0.75\textlength}
  \makebox[\textlength][s]{$#2$} \hspace{-.55\textlength}
  \hspace{-\overlinelength}\hspace{#1\overlinelength}
  \overline{\makebox[\overlinelength][s]{\vphantom{$#2$}}}
  \hspace{-#1\overlinelength}\hspace{.55\textlength}}

\DeclareMathOperator{\diag}{diag}
\DeclareMathOperator{\tr}{tr}

\newcommand{\VEV}[1]{\left\langle #1\right\rangle}
\newcommand{\VEVsmall}[1]{\langle #1\rangle}

\begin{document}

\begin{center}
  {\Large\bfseries Perturbative SO(10) GUT and the Minimal Higgs
    Sector}
  \\[9pt]
  {\large S\"oren Wiesenfeldt and Scott Willenbrock}
  \\[9pt]
  {\normalsize {\slshape
      \begin{minipage}{.8\linewidth}
        \centering Department of Physics, University of Illinois at
        Urbana-Champaign,
        \\
        1110 West Green Street, Urbana, IL 61801, USA
      \end{minipage}
    }}
\end{center}

\begin{abstract}
  \noindent
  The breaking of SO(10) to $\text{SU(3)}_C \times
  \text{U(1)}_\text{EM}$ can be accomplished by just four Higgs
  fields: the symmetric rank-two tensor, $S(\mathsf{54})$; a pair of
  spinors, $C(\mathsf{16})$ and $\ovl{C}(\ovl{\mathsf{16}})$; and a
  vector, $T(\mathsf{10})$.  This setup is also able to generate
  realistic fermion masses.  The heavy color triplets in the vector
  and spinor fields mediate proton decay via dimension-five operators.
  The experimental bounds on proton decay constrain the structure and
  size of the Yukawa operators.
\end{abstract}

\noindent
SO(10) \cite{so10} is arguably the most natural grand-unified theory
(GUT): both the standard model (SM) gauge and matter fields are
unified, introducing only one additional matter particle, the
right-handed neutrino.  It is an anomaly-free theory and therefore
explains the intricate cancellation of the anomalies in the standard
model \cite{Georgi:1972bb}.  Moreover, it contains $B-L$ as a local
symmetry, where $B$ and $L$ are baryon and lepton number,
respectively; the breaking of $B-L$ naturally provides light neutrino
masses via the seesaw mechanism.

Despite these attractive features, the breaking of the GUT symmetry
has remained a problem in model building.  Generally, two avenues have
been pursued.  One makes use of large representations,
$\Phi(\mathsf{210})$, $\Sigma(\mathsf{126})$,
$\ovl{\Sigma}(\ovl{\mathsf{126}})$, and $T(\mathsf{10})$ \cite{large}.
This approach has the advantage of a fully renormalizable
superpotential and automatic $R$ parity.  The unified gauge coupling,
however, diverges just above the GUT scale, indicating that new
physics must enter.  Since this new physics is close to the GUT scale,
it potentially has a large effect on the predictions of the model.  In
addition, this scenario is insufficient to reproduce the fermion mass
spectrum, have successful gauge unification, and fulfill the proton
decay constraints at the same time \cite{Melfo:2007sr}.  Hence, a
realistic renormalizable model requires at least another Higgs field
\cite{large-new}.

Models with only small representations remain perturbative up to the
Planck scale \cite{Chang:2004pb} and also have the potential to arise
from string theory.  They introduce a moderate number of new fields,
yielding only small threshold corrections at $M_\text{GUT}$.  In the
supersymmetric version of this scenario, higher-dimensional operators,
suppressed by powers of a more fundamental scale $M$ (such as the Planck
scale, $M_\text{P} = \left(8\pi G_N\right)^{-1/2} = 2\cdot 10^{18}$~GeV,
or the string scale in the weakly coupled heterotic string, $M_\text{S}
\approx 5\cdot 10^{17}$~GeV), are essential to achieve the breaking to the SM
group, $\text{G}_\text{SM} = \text{SU(3)}_C \times \text{SU(2)}_L \times
\text{U(1)}_Y$ \cite{Barr:1997hq}.  Moreover, these operators play an
important role in fermion masses and mixings \cite{bpw,so10-small}.
First, they generate Majorana masses for the right-handed neutrinos in
the desired range, $M_\text{GUT}^2/M\sim 10^{14}$ GeV.  Thus neutrino
masses require the vacuum expectation value (vev) of the $B-L$ breaking
field to be of the order of $M_\text{GUT}$.  Second, they naturally
explain why certain relations, such as the bottom-tau unification, are
only approximately realized.  Finally, they alter the couplings of the
matter fields to the Higgs color triplets relative to the weak doublets,
such that the proton decay rate via dimension-five operators can be
significantly reduced \cite{Emmanuel-Costa:2003pu}.

Models with small representations usually use the antisymmetric
second-rank tensor, $A(\mathsf{45})$, together with a pair of spinors,
$C(\mathsf{16})$ and $\ovl{C}(\ovl{\mathsf{16}})$.  In order to give
GUT-scale masses to the color-triplet Higgs fields, one must introduce
two ten-dimensional Higgs fields because the term $TA\,T$ vanishes due
to the antisymmetry of $A$.  This setup can implement the
Dimopoulos-Wilczek mechanism \cite{dw,Babu:1993we}, yielding automatic
mass splitting of the (light) Higgs doublets and (heavy) color
triplets.  However, the realization of this mechanism requires a
second pair of spinorial Higgs fields and an extensive set of global
symmetries \cite{Barr:1997hq}.  Thus it is important to investigate
alternative scenarios.\footnote{One attempt has been to realize the
  GUT gauge symmetry in more than four space-time dimension and use
  GUT-symmetry breaking boundary conditions on an orbifold for the
  breaking to the SM.  This scenario yields doublet-triplet splitting
  and avoids the dangerous dimension-five proton-decay operators
  \cite{orbifold}.}

Although the breaking of SO(10) to the standard model by the symmetric
rank-two tensor, $S(\mathsf{54})$, and a spinorial, $B-L$ breaking
field is a standard textbook example \cite{books}, a realistic
supersymmetric model has never been worked out.\footnote{The Higgs
  sector for this setup was considered for the non-supersymmetric case
  \cite{non-susy} and also for the supersymmetric scenario, but with
  two ten-dimensional fields \cite{Zhao:1981me}.  In several
  supersymmetric models, $S$ is used as an additional field to achieve
  the symmetry breaking \cite{45-54}.}  The reason might be twofold.
First, this scenario does not allow for the Dimopoulos-Wilczek
mechanism, so we have to accept fine-tuning to have one pair of Higgs
doublets light.  Second, $S(\mathsf{54})$ breaks SO(10) to the
Pati-Salam group, $\text{G}_\text{PS} = \text{SU(4)}_C \times
\text{SU(2)}_L \times \text{SU(2)}_R$.  Hence, in contrast to $A$, it
does not break $\text{SU(4)}_C$ and therefore preserves the
unification of down-quark and charged-lepton masses.  Thus, with only
small Higgs representations, it is more involved to generate realistic
quark and lepton masses.

The use of $S(\mathsf{54})$, however, has advantages over both models
described above.  In contrast to the model with large representations,
the gauge coupling remains perturbative up to $M_\text{P}$ and unlike
the scenario with $A(\mathsf{45})$, it requires only one
ten-dimensional Higgs field for the electroweak symmetry
breaking.\footnote{Recently, it was shown that SO(10) can be broken to
  $\text{G}_\text{SM}$ by a single pair of vector-spinors,
  $\Upsilon_{(+)}(\mathsf{144}) + \Upsilon_{(-)}(\ovl{\mathsf{144}})$
  \cite{Babu:2005gx}; however, in order to generate realistic fermion
  masses, one needs to introduce additional matter fields, such as
  $\text{10}_M$ and $\text{45}_M$ \cite{Babu:2006rp}.  We will not
  consider this approach in this letter.}  In this letter we
demonstrate that SO(10) can be broken to $\text{SU(3)}_C \times
\text{U(1)}_\text{EM}$ by a set of four Higgs fields:
$S(\mathsf{54})$, $C(\mathsf{16})$, $\ovl{C}(\ovl{\mathsf{16}})$, and
$T(\mathsf{10})$.  Analyzing the higher-dimensional operators, we show
that realistic fermion masses and mixings can be generated and a
sufficiently low proton decay rate can be obtained.

\paragraph{Breaking of SO(10).}
The SO(10) symmetry is broken by the vevs of $S(\mathsf{54})$,
$C(\mathsf{16})$, and $\ovl{C}(\ovl{\mathsf{16}})$ \cite{non-susy}.
The superpotential can be split into three parts,
\begin{align}
  W & = W_S + W_C + W_{SC} \;.
  \label{eq:superpotential}
\end{align}
The first part yields the breaking to $\text{SO(6)} \times
\text{SO(4)} \simeq \text{G}_\text{PS}$,
\begin{align}
  \label{eq:superpotential-S}
  W_S & = \tfrac{1}{2}\, M_S \tr S^2 + \tfrac{1}{3} \lambda_S \tr S^3
  \;, \qquad \VEV{S} = v_s \diag(2,2,2,2,2,2;-3,-3,-3,-3) \;, \qquad
  v_s = \frac{M_S}{\lambda_S} \;.
\end{align}
$\VEV{S}$ is chosen such that the first six entries preserve
$\text{SO(6)} \simeq \text{SU(4)}_C$, whereas the last preserve
$\text{SO(4)} \simeq \text{SU(2)}_L \times \text{SU(2)}_R$.
The second part of the superpotential describes the breaking of SO(10)
to SU(5) by the vevs of the spinorial representations.  It requires
the inclusion of one dimension-five operator,\footnote{Alternatively,
  we may introduce a singlet $X$ such that
  \begin{align*}
    W_C & = X \left( \lambda_C C \ovl{C} - M_C^2 \right) , \qquad
    \frac{\partial W_C}{\partial X} = \lambda_C C \ovl{C} - M_C^2 \
    \Rightarrow\ \VEV{C\ovl{C}} = \frac{M_C^2}{\lambda_C} \ , \qquad
    \frac{\partial W_C}{\partial C} = \lambda_C X \ovl{C} \
    \Rightarrow \VEV{X} = 0 \;,
  \end{align*}
  avoiding the dimension-five operator.  The price we pay is to
  introduce a singlet which in general can couple to the other
  fields as well.}
\begin{align}
  \label{eq:superpotential-C}
  W_C & = M_C\, C \ovl{C} + \frac{\lambda_C}{2M} \left( C \ovl{C}
  \right)^2 , \qquad \VEV{C\ovl{C}} = -\frac{M_C M}{\lambda_C} \equiv
  v_c^2 \;.
\end{align}
Together, these vevs break SO(10) to the intersection of SU(5) and
$\text{G}_\text{PS}$, namely the standard model group
$\text{G}_\text{SM}$.

At the renormalizable level, the Higgs fields do not couple in the
potential.  Therefore we have two independent global SO(10) symmetries
and the total number of Goldstones is $\left(45-24\right) +
\left(45-21\right) = 45$.  Only $45-12=33$ true Goldstones are eaten
by gauge bosons, so there are 12 pseudo-Goldstones.  These are a
vectorial pair of quark-doublet fields,
$Q(3,2,\frac{1}{6})+\ovl{Q}(\bar{3},2,-\frac{1}{6})$, contained both
in $C + \ovl{C}$ and in $S$.  This extra global symmetry, however, is
accidental, and is violated by the non-renormalizable interaction term
\begin{wrapfigure}{r}{.42\linewidth}
  \centering
  \vspace*{-9pt}
  \captionsetup{margin=20pt}
  \caption{\slshape The running of the gauge couplings at 1-loop in the
    MSSM (dashed) and with the inclusion of the pseudo-Goldstones at $3
    \cdot 10^{14}$ GeV (solid).  The MSSM couplings unify at around $2
    \cdot 10^{16}$ GeV, while the inclusion of the pseudo-Goldstones at
    a lower scale disrupts the unification.  \vspace*{27pt}
    \label{fig:running}
  }%
  \includegraphics[width=\linewidth]{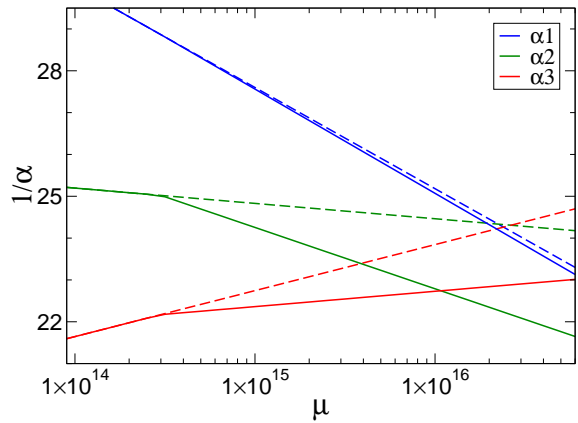}
\end{wrapfigure}
between the different Higgs fields,
\begin{align}
  W_{SC} & = \frac{\xi_C}{2M}\, C \ovl{C} S^2 .
  \label{eq:superpotential-SC}
\end{align}
The pseudo-Goldstones acquire masses of order $v_s v_c/M$.  Their
presence changes the 1-loop coefficients of the $\beta$-function of the
gauge couplings $\left(\alpha_1,\alpha_2,\alpha_3\right)$ by $\Delta
b=\left(\frac{1}{5},3,2\right)$ such that
$b=\left(\frac{34}{5},4,-1\right)$ \cite{Barger:2006fm}.  Hence, they
mostly modify the running of $\alpha_2$ and $\alpha_3$.
Fig.~\ref{fig:running} shows the impact of the pseudo-Goldstones, at a
mass of $3 \cdot 10^{14}$ GeV, on the running of the gauge couplings in
the MSSM at one loop.  Since the gauge coupling unification is upset, it
is clear that the pseudo-Goldstone masses must be close to the GUT scale
in order to preserve gauge coupling unification.  This can be achieved
provided that $v_c > v_s$; if $\lambda_C \sim \frac{M_C}{M}$, the
pseudo-Goldstones become as heavy as $v_s$.  In this scenario, SO(10) is
first broken to SU(5) at $v_c$, and then to the SM at $v_s$.
A drawback to this scenario is that $v_c \sim M$, so even
higher-dimensional operators obtained by adding powers of
$\left(C\ovl{C}\right)/M^2$, $\left(C/M\right)^4$, and $(\ovl{C}/M)^4$
are not necessarily suppressed.

\medskip
 
The renormalizable part of the superpotential for the electroweak
symmetry breaking reads
\begin{align}
  \label{eq:superpotential-T}
  W_T & = \tfrac{1}{2}\, M_T\, T^2 + \tfrac{1}{2}\, \lambda_T T S T +
  \xi_T\, C C T + \xi_T^\prime\, \ovl{C} \ovl{C} T \; .
\end{align}
The mass matrix for the weak doublets, $H_u(T)$, $H_d(T)$,
$\widetilde{H}_u(\bar{C})$, and $\widetilde{H}_d(C)$, is such that
without fine-tuning, all doublets have GUT-scale masses,
\begin{align}
  \label{eq:m2-matrix}
  \begin{pmatrix}
    H_u & \widetilde{H}_u
  \end{pmatrix}
  \begin{pmatrix}
    M_T - 3 \lambda_T v_s & \xi_T v_c \cr \xi_T^\prime v_c & M_C +
    \frac{\lambda_C}{M}\, v_c
  \end{pmatrix}
  \begin{pmatrix}
    H_d \cr \widetilde{H}_d
  \end{pmatrix}
  .
\end{align}
This is the well-known doublet-triplet splitting problem.  In order to
arrange for two light doublets, we have to impose that the determinant
of this mass matrix vanishes, up to weak-scale terms.  As a result, the
two light doublets are combinations of the four doublets in $T$, $C$,
and $\ovl{C}$, and all doublets acquire weak-scale vevs.\footnote{This
  is similar to the scenario with large representations, where the
  doublets mix through the vev of $\Phi(\mathsf{210})$ and the light
  doublets are mixtures of those in $T(\mathsf{10})$,
  $\Sigma(\mathsf{126})$, $\ovl{\Sigma}(\ovl{\mathsf{126}})$, and
  $\Phi(\mathsf{210})$ \cite{large}.}  This is crucial in order to
derive a realistic fermion mass spectrum, as we will discuss in the next
section \cite{bpw,so10-small}.
We will denote the SU(5)-singlet and SU(2)-doublet vevs of $C$ as
$\VEVsmall{C^\text{GUT}}$ and $\VEVsmall{C^\text{EW}}$ and similarly the
vevs of $\ovl{C}$.

For later purposes, note that due to the vev of $S$ in
Eq.~(\ref{eq:superpotential-S}), the color triplets of any
ten-dimensional representation occupy the first six entries, whereas
the last four are for the weak doublets.  For $T$, e.g., we may write
\mbox{$T=\left(H_C,\ovl{H}_C;H_u,H_d\right)$}, where $H_C$ and
$\ovl{H}_C$ are the proton-decay-mediating color triplets.

\paragraph{Fermion Masses.}
There is only one renormalizable operator that generates masses for
the matter fields $\mathsf{16}_i$, $i=1,2,3$,
\begin{align}
  \label{eq:yukawa-dim4}
  W_Y^{(4)} & = h_T^{ij}\, \mathsf{16}_i \mathsf{16}_j T ;
\end{align}
this term predicts Yukawa unification at $M_\text{GUT}$,
$h_u=h_d=h_e=h_\nu^\text{D}$.  To have a realistic mass pattern, we
need to consider higher-dimensional operators.  At dimension five, we
have
\begin{align}
  \label{eq:yukawa-dim5}
  W_Y^{(5)} & = \frac{1}{M} \left[ h_S^{ij}\, \mathsf{16}_i
    \mathsf{16}_j T S + h_C^{ij}\, \mathsf{16}_i \mathsf{16}_j C C +
    h_{\bar{C}}^{ij}\, \mathsf{16}_i \mathsf{16}_j \ovl{C} \ovl{C}
  \right] .
\end{align}
These operators can be generated by integrating out heavy fields in
various SO(10) representations, as indicated in Table~\ref{tb:yukawa}.

The first term in Eq.~(\ref{eq:yukawa-dim5}) contributes equally to the
mass matrices of quarks and leptons since the vev of $S$ conserves
$\text{SU(4)}_C$.  This is different from the scenario with the
antisymmetric representation, which breaks SO(10) to the left-right
symmetric group, $\text{SU(3)}_C \times \text{SU(2)}_L \times
\text{SU(2)}_R \times \text{U(1)}_{B-L}$.  The second term contributes
(in equal measure) to the masses of down quarks and charged leptons,
whereas the last term generates Dirac masses for up quarks and neutrinos
as well as Majorana masses for the right-handed neutrinos
\cite{Sayre:2006wf}.  These terms allow for a non-trivial CKM-matrix and
degrade the Yukawa relations to $h_d=h_e$.  The relation
$h_u=h_\nu^\text{D}$ is violated by the operators
\mbox{$\left[\mathsf{16}_i \ovl{C}\right] \left[ \mathsf{16}_j \ovl{C}
  \right]$}, obtained by integrating out heavy fields in either the
singlet or adjoint representation, as shown in Table~\ref{tb:yukawa}
\cite{Babu:2000ei}.

In order to alter the unification of down quark and charged lepton
masses, we have to go further and consider terms containing both $S$
and $C$ fields.  The lowest operator is of dimension six,
\begin{align}
  \label{eq:yukawa-dim6}
  W_Y^{(6)} & = \frac{h_{SC}^{ij}}{M^2}\, \mathsf{16}_i \mathsf{16}_j
  C C S \;,
\end{align}
suppressed by $v_c v_s/M^2$.  For $M=M_\text{P}$, this suppression
factor is of similar order of magnitude as the ratio of the strange
quark mass and the weak scale, namely $10^{-3}$.  Thus this term is
large enough to account for the difference of strange-quark and muon
masses.  We will now show that this term does indeed violate the
equality of down-quark and charged-lepton masses.

Actually, there are two distinct operators in
Eq.~(\ref{eq:yukawa-dim6}).  To see that, let us recall that the Higgs
field $\mathsf{120}$ contains two pairs of weak doublets.
The first pair is that of the SU(5)-fields $\text{5} + \ovl{\text{5}}$
of SU(5) and $\left(1,2,2\right)$ of $\text{G}_\text{PS}$, and it
couples equally to down quarks and charged fermions.  (The same applies
to the pair of Higgs doublets in $T$.)  The second pair of doublets is
contained in $\text{45} + \ovl{\text{45}}$ of SU(5) and $(15,2,2)$ of
$\text{G}_\text{PS}$.  Here the couplings to the fermion fields pick up
a $B-L$ factor, modifying the Yukawa unification \cite{Ross:1985ai}.

In Eq.~(\ref{eq:yukawa-dim6}), the fermions effectively couple to
$\mathsf{10}$, $\mathsf{120}$, or $\ovl{\mathsf{126}}$ fields in the
$\mathsf{16} \times \mathsf{16} \times \mathsf{54}$ decomposition.  We
expect two types of couplings to be present, one that yields Yukawa
unification, and one that violates it.  Let us demonstrate that the
operator in Eq.~(\ref{eq:yukawa-dim6}) can indeed modify the Yukawa
unification by integrating out heavy
$\mathsf{10}_M=\left(D,D^c;L,L^c\right)_M$ fields.
In this case, the two types of couplings are \mbox{$h_{SC}^{(A)\,ij}
  \left[ \mathsf{16}_i \mathsf{16}_j \right]_\mathsf{10} S \left[ C C
  \right]_\mathsf{10}$} and \mbox{$h_{SC}^{(B)\,ij} \left[ \mathsf{16}_i
    C \right]_\mathsf{10} S \left[ \mathsf{16}_j C
  \right]_\mathsf{10}$}.  The first operator gives equal contributions
to down quark and charged lepton masses, so let us study the second in
detail.

The coupling of the matter fields $\mathsf{16}_i$ and $\mathsf{10}_M$ to
$C^\text{EW}$ (where $\widetilde{H}_d(C)$ acquires its vev) yields
\begin{align}
  \label{eq:yukawa-spoil-ew}
  \mathsf{16}_i \mathsf{10}_M C^\text{EW} \ni \left( d_i D^c_M + e^c_i
    L_M + \nu^c_i L^c_M \right) \widetilde{H}_d ,
\end{align}
in terms of SO(10) and SM fields.  The coupling to $C^\text{GUT}$ (where
the SM singlet component, $N(C)$, acquires its vev) gives
\begin{align}
  \label{eq:yukawa-spoil-gut}
  \mathsf{16}_j \mathsf{10}_M C^\text{GUT} \ni \left( d^c_j D_M + L_j
    L^c_M \right) N .
\end{align}
As noted above, the color triplets of the SO(10) vector live in the
first six entries, the doublets in the last four.  Then we may write
\mbox{$\left[ \mathsf{16} \VEV{C^\text{EW}} \right] =
  \left(d;e^c,0\right) v_d$} and $\left[ 16 \VEV{C^\text{GUT}} \right]
= \left(d^c;L\right) v_c$, where $\VEVsmall{\widetilde{H}_d}=v_d$.
Now it is straightforward to see that this operator indeed violates
the Yukawa unification,
\begin{align}
  \label{eq:yukawa-spoil}
  \frac{h_{SC}^{(B)\,ij}}{M^2} \left[ \left(d;e^c,0\right)_i
    \VEVsmall{C^\text{EW}} \vphantom{D^j_j} \right] \VEV{S} \left[
    \left(d^c;L\right)_j \VEVsmall{C^\text{GUT}} \right] =
  h_{SC}^{(B)\,ij} \left( 2\, d_i d^c_j - 3\, e^c_i e_j \right) v_d
  \frac{v_c v_s}{M^2} \;.
\end{align}

\begin{table}
  \centering
  \renewcommand{\arraystretch}{1.2}
  \captionsetup{margin=13pt,position=top}
  \caption{\slshape Yukawa operators up to dimension six and their relative contributions
    to the fermion mass matrices (left columns) and baryon and lepton
    number violating couplings (right columns).  For various
    operators, there exist several options to contract the indices.
    For $\mathsf{16}_i \mathsf{16}_j \ovl{C} \ovl{C}$, these lead to
    different results as indicated by the heavy field in brackets
    integrated out to generate the operator.
    The label $(A)$ indicates the grouping \mbox{$\left[
        \mathsf{16}_i \mathsf{16}_j \right] S \left[ C C \right]$},
    while $(B)$ indicates \mbox{$\left[ \mathsf{16}_i C \right] S \left[
        \mathsf{16}_j C \right]$}.
  }
  \label{tb:yukawa}
  \begin{tabular}{llr|dddd|dddd}
    \hline
    operator & & & \multicolumn{1}{c}{$Y_u$} &
    \multicolumn{1}{c}{$Y_d$} & \multicolumn{1}{c}{$Y_e$} &
    \multicolumn{1}{c|}{$Y_\nu^\text{D}$} &
    \multicolumn{1}{c}{$Y_{qq}$} & \multicolumn{1}{c}{$Y_{ql}$} &
    \multicolumn{1}{c}{$Y_{ud}$} & \multicolumn{1}{c}{$Y_{ue}$} 
    \\
    \hline
    $h_T^{ij}\,\mathsf{16}_i \mathsf{16}_j T$ 
    & & & 1 & 1 & 1 & 1 & 1 & 1 & 1 & 1
    \\
    \hline
    $h_S^{ij}\,\mathsf{16}_i \mathsf{16}_j T S$ 
    & & & -3 & -3 & -3 & -3 & 2 & 2 & 2 & 2
    \\
    $h_C^{ij}\,\mathsf{16}_i \mathsf{16}_j C C$ 
    & & & - & 1 & 1 & - & - & 1 & 1 & - 
    \\
    $h_{\bar{C}}^{ij}\,\mathsf{16}_i \mathsf{16}_j \ovl{C} \ovl{C}$
    & & [1] & - & - & - & 1 & - & - & - & -
    \\
    & & [10] & 1 & - & - & 1 & 1 & - & - & 1
    \\
    & & [45] & 8 & - & - & 3 & 8 & - & - & 8
    \\
    \hline
    $h_{SS}^{ij}\,\mathsf{16}_i \mathsf{16}_j T S S$ 
    & & & 9 & 9 & 9 & 9 & 4 & 4 & 4 & 4
    \\
    $h_{SC}^{ij}\,\mathsf{16}_i \mathsf{16}_j C C S$ 
    & $(A)$ & & - & -3 & -3 & - & - & 2 & 2 & -
    \\
    & $(B)$ & & - & 2 & -3 & - & - & -3 & 2 & -
    \\
    $h_{S\bar{C}}\,\mathsf{16}_i \mathsf{16}_j \ovl{C} \ovl{C} S$ 
    & & & -3 & - & - & -3 & 2 & - & - & 2
    \\
    \hline
  \end{tabular}
\end{table}

The various operators up to dimension six and their contributions to
the quark and lepton mass matrices are listed in the left part of
Table~\ref{tb:yukawa}.  Remarkably, the operator $\mathsf{16}_i
\mathsf{16}_j \ovl{C} \ovl{C} S$ contributes in equal measure to $h_u$
and $h_\nu^\text{D}$, in contrast to the dimension-five operator
$\mathsf{16}_i \mathsf{16}_j \ovl{C} \ovl{C}$.
From Table~\ref{tb:yukawa}, we read off the relations
\begin{align}
  \label{eq:mass-relations}
  Y_\nu^\text{D} - Y_u & = h_{\bar{C}}^{[1]} - 5\, h_{\bar{C}}^{[45]}
  , \qquad Y_e - Y_d = 5\, h_{SC}^{(B)} \;.
\end{align}

\paragraph{Proton Decay.}
The couplings of the fermions to color-triplet Higgs fields give rise
to the proton decay operators of mass-dimension five \cite{dim5op}
with
\begin{align}
  \Gamma & \propto \left| \frac{C_5}{M_{H_C}} \right|^2 , \qquad C_5^L
  = Y_{qq} Y_{ql} \;, \qquad C_5^R = Y_{ud} Y_{ue} \;,
\end{align}
where $\Gamma$ is the decay rate, $M_{H_C}$ is the mass of the color
triplets, and the baryon and lepton number violating couplings are
denoted as
\begin{align}
  \label{eq:pdecay-cplg}
  \left( \tfrac{1}{2}\,Y_{qq}^{ij}\, Q_i Q_j + Y_{ue}^{ij}\, u^c_i
    e^c_j \right) H_C + \left( Y_{ql}^{ij}\, Q_i L_j + Y_{ud}^{ij}\,
    u^c_i d^c_j \right) \ovl{H}_C \, .
\end{align}
\begin{wrapfigure}{r}{.37\linewidth}
  \flushright
  \scalebox{.9}{
    \begin{picture}(160,35)(20,0)
      \ArrowLine(30,50)(80,30)   \Text(22,53)[]{$u$}
      \ArrowLine(30,10)(80,30)   \Text(22,12)[t]{$d$}
      \Vertex(80,30)2            \Text(80,17)[t]{$C_5^L=Y_{qq}Y_{ql}$}
      \DashLine(80,30)(130,50)5  \Text(117,60)[t]{$\widetilde l$}
      \DashLine(80,30)(130,10)5  \Text(117,27)[t]{$\widetilde q$}
      \ArrowLine(160,50)(130,50) \Text(172,52)[t]{$\nu$}
      \Vertex(130,50)1           
      \Line(130,10)(130,50)
      \Photon(130,10)(130,50){2}{6}
      \Text(142,30)[]{$\widetilde{w}^\pm$} 
      \Vertex(130,10)1           
      \ArrowLine(160,10)(130,10) \Text(169,12)[t]{$s$}
    \end{picture}
  }
  \scalebox{.9}{
    \begin{picture}(160,80)(20,10)
      \ArrowLine(30,50)(80,30)   \Text(22,53)[]{$u^c$}
      \ArrowLine(30,10)(80,30)   \Text(22,12)[t]{$d^c$}
      \Vertex(80,30)2            \Text(80,17)[t]{$C_5^R=Y_{ud}Y_{ue}$}
      \DashLine(80,30)(130,50)5  \Text(117,50)[]{$\widetilde e^c$}
      \DashLine(80,30)(130,10)5  \Text(117,25)[]{$\widetilde u^c$}
      \ArrowLine(160,50)(130,50) \Text(172,52)[t]{$\nu$}
      \Vertex(130,50)1           
      \ArrowLine(130,30)(130,50)
      \ArrowLine(130,30)(130,10) \Vertex(130,30)1
      \Text(142,30)[]{$\widetilde{h}^\pm$} 
      \Vertex(130,10)1           
      \ArrowLine(160,10)(130,10) \Text(169,12)[t]{$s$}
    \end{picture}
  }
\end{wrapfigure}
Decay diagrams $p \to K^+ \bar\nu$ via the two distinct operators
$QQQL$ and $u^c d^c u^c e^c$ are sketched in the adjoining figure.

The determination of the baryon and lepton number violating couplings
is important for the calculation of the decay amplitude.  In SU(5),
the impact of higher-dimensional Yukawa operators on these couplings,
relative to the mass terms, is sufficient to reduce the decay rate by
several orders of magnitude and make it consistent with the
experimental upper bound \cite{Emmanuel-Costa:2003pu}.

The dimension-five operators are generated by integrating out the
heavy color-triplet Higgs fields.  In addition to the standard
couplings to $H_C(T)$ and $\ovl{H}_C(T)$, we have those to
$\widetilde{H}_C(\ovl{C})$ and $\widetilde{\ovl{H}}_C(C)$, via the
higher-dimensional operators \cite{bpw}.
The coefficients are listed in the right part of
Table~\ref{tb:yukawa}; note that the first two rows express the
couplings to $H_C$ and $\ovl{H}_C$, the remaining rows those to
$\widetilde{H}_C$ and $\widetilde{\ovl{H}}_C$.

Due to the two pairs of color triplets, we cannot simply read off the
relations between the baryon-number-violating couplings and the mass
matrices, in contrast to SU(5).  We find, however, relations among the
four different couplings, namely
\begin{align}
  Y_{qq} & = Y_{ue}
\end{align}
and the SU(5) relation \cite{Emmanuel-Costa:2003pu,Bajc:2002pg}
\begin{align}
  Y_{ud} - Y_{ql} & = Y_d - Y_e \,.
\end{align}
A detailed study of the various decay modes requires a numerical
analysis of the fermion masses and mixings, which is beyond the scope
of this letter.  Note that the new decay operators due to the color
triplets in $C$ and $\ovl{C}$ can change the branching ratios
significantly \cite{bpw}.  Na\"ive choices such as $Y_u=Y_{qq}$ yield
a decay rate which could be in conflict with the experimental bounds
\cite{Kobayashi:2005pe}, provided that the color triplets have
GUT-scale masses.  A study along the lines of
Ref.~\cite{Emmanuel-Costa:2003pu}, however, would provide means to
control the total decay rate.  The experimental bounds on proton
decay, together with the observed pattern of fermion masses and
mixings will constrain the structure and size of the various
couplings.

\paragraph{Concluding Remarks.}
We have shown that SO(10) can be broken to $\text{SU(3)}_C \times
\text{U(1)}_\text{EM}$ by a set of four Higgs fields:
$S(\mathsf{54})$, $C(\mathsf{16})$, $\ovl{C}(\ovl{\mathsf{16}})$, and
$T(\mathsf{10})$.  This raises the question of whether this scenario
is the minimal set of Higgs fields capable of breaking SO(10) to
$\text{SU(3)}_C \times \text{U(1)}_\text{EM}$.  If we replace the
symmetric tensor $S(\mathsf{54})$ by the antisymmetric tensor,
$A(\mathsf{45})$, will this system work as well?

In order for $A$ to acquire a vev, we need a quartic coupling, such as
$\frac{1}{M} \tr A^4$.  Even if it acquires a vev in the $B-L$
direction, the Dimopoulos-Wilczek form is destabilized when $A$ couples
to the spinors via the operator $C A \ovl{C}$ and the $I_{3R}$
component gets a vev as well \cite{Barr:1997hq,Babu:1993we}. (If $A$
acquires a vev in the hypercharge direction, both the $B-L$ and
$I_{3R}$ components of $A$ have a vev from the beginning.)
Moreover, this coupling generates a splitting among the electroweak
doublets and color triplets in $C$ and $\ovl{C}$.  Hence, although
there is not a renormalizable coupling of $A$ to $T$, the
doublet-triplet splitting emerges and fine-tuning allows to arrange
for two light weak doublets, while the color triplets are heavy.
Therefore this scenario is capable of breaking SO(10) to
$\text{SU(3)}_C \times \text{U(1)}_\text{EM}$ as well.

Comparing the two scenarios with either $A(\mathsf{45})$ or
$S(\mathsf{54})$, we notice that in order to reproduce fermion masses,
we only need operators of dimension five in the former case.  In the
latter scenario, however, we have seen that the dimension-six operators
are necessary, and have a significant impact.  Dimension-six operators,
although not necessary, could have a significant impact in the models
with $A(\mathsf{45})$ as well.  Furthermore, looking at
Table~\ref{tb:yukawa}, we note that several operators contribute
identically to the fermion mass matrices.  This reduces the number of
free parameters in these matrices.  Hence, using $S(\mathsf{54})$ is a
promising alternative to the already established scenarios.


\smallskip
  
We are grateful for valuable discussions with K.~Babu and to
C.~Albright for comments on the manuscript.
This work was supported in part by the U.~S.~Department of Energy
under contract No.~DE-FG02-91ER40677.

\newpage

{\small
  
}%

\end{document}